\documentclass[10pt,aps,prf,showpacs,floats,onecolumn,superscriptaddress,longbibliography]{revtex4-2}

\usepackage{comment}
\DeclareUnicodeCharacter{2009}{\,}
\UseRawInputEncoding

\usepackage{lipsum}
\usepackage{mathrsfs}
\usepackage{bm,amsbsy,amssymb,amsmath}
\usepackage{graphics,graphicx,dcolumn,epic,eepic,float,tabularx}
\usepackage{multirow,rotate,rotating,color}
\usepackage[utf8]{inputenc}
\usepackage{relsize}

\usepackage{xcolor}
\definecolor{tuered}{RGB}{214,0,74}
\definecolor{tueblue}{RGB}{0,102,204}

\usepackage{hyperref}

\begin{document}
\title{Coalescence-induced alignment of anisotropic particles in drying sessile droplets}

\author{Johannes  Sch\"ottner}
\email{j.schoettner@fz-juelich.de}
\affiliation{Helmholtz Institute Erlangen-N\"urnberg for Renewable Energy (IET-2), Forschungszentrum J\"ulich, Cauerstra{\ss}e 1, 91058 Erlangen, Germany}

\author{Qingguang Xie}
\email{q.xie@fz-juelich.de}
\affiliation{Helmholtz Institute Erlangen-N\"urnberg for Renewable Energy (IET-2), Forschungszentrum J\"ulich, Cauerstra{\ss}e 1, 91058 Erlangen, Germany}

\author{Jens Harting}
\email{j.harting@fz-juelich.de}
\affiliation{Helmholtz Institute Erlangen-N\"urnberg for Renewable Energy (IET-2), Forschungszentrum J\"ulich, Cauerstra{\ss}e 1, 91058 Erlangen, Germany}
\affiliation{Department of Chemical and Biological Engineering and Department of Physics, Friedrich-Alexander-Universit\"at Erlangen-N\"urnberg, Cauerstra{\ss}e 1, 91058 Erlangen, Germany}


\begin{abstract}

\vspace{10px}

Alignment of anisotropic particles strongly governs the functional properties of printed materials, yet most studies have focused on particle alignment in single evaporating droplets. In droplet-based printing, however, neighboring droplets can coalesce, generating rapid capillary flows that redistribute material and markedly affect the final morphology. 
Here, we use mesoscale simulations to investigate how droplet coalescence and subsequent evaporation jointly determine alignment and redistribution in sessile droplets with different contact angles and volumes. During the early stages of coalescence, the mean nematic order along the coalescence direction increases for all combinations of contact-angle and volume asymmetries of the droplets. At later times, the mean nematic order either continues to increase or decreases, depending on the droplet geometry. We derive a geometric scaling based on curvature and volume asymmetry and show that the simulation results collapse onto a master curve, identifying an effective geometric asymmetry parameter that governs the mean nematic order at the end of coalescence. 
During evaporation, the contact angle strongly influences how the coalescence-induced orientational structure is transferred to the final deposit. For small contact angles, the contact line remains pinned for a longer duration, better preserving alignment. In contrast, larger contact angles promote contact-line motion, which weakens alignment, as reflected by a reduced mean nematic order, while simultaneously generating stronger concentration gradients in the final deposit.

\end{abstract}  

\maketitle

\section{Introduction}

The controlled alignment of anisotropic particles is crucial for printed functional materials, as the resulting particle arrangement governs network connectivity, anisotropy, crystallization, and thus the functional response of the deposited structure~\cite{Dong2025, Cuiling2025}. In droplet-based printing, capillary flows during droplet drying, together with interfacial confinement, strongly influence particle alignment. Most studies have therefore focused on particle alignment in single evaporating droplets~\cite{Qingchang2018, Dugyala2015}. However, in practical printing processes, neighboring droplets may coalesce and subsequently dry, and both processes are expected to significantly affect the morphology of the deposit.

Coalescence has been studied extensively in related systems, with particular focus on bridge growth, relaxation dynamics, and asymptotic scaling laws for different coalescence regimes and contact-angle limits~\cite{Hack2020,kaneelil2026, Thomas2023, eggers2025}. Sessile droplet coalescence has also been investigated for complex fluids, including polymeric, viscoelastic, and yield-stress droplets and inks~\cite{Varma2021, varma2022, Sivasankar2022, arbabi2023, franca2026}. Furthermore, practical deposition often involves asymmetry between neighboring droplets. They may differ in contact angle, volume, composition, surface tension, viscosity, surfactant concentration, or particle loading. Such differences can arise from different formulations, local evaporation histories, substrate heterogeneity, or particle-induced pinning~\cite{Farrokhi2025}. Asymmetric coalescence can generate directional flow, delayed relaxation, incomplete mixing, or even flow reversal during the approach to the final equilibrium shape~\cite{hernandezsanchez2012, xu2023, Pawar2019, Hack2021, Xin2012}. Mixing of different liquids or polymeric materials adds another level of complexity, because composition-dependent viscosity, surface tension, and internal flow can influence both coalescence and the final material distribution~\cite{SykesLangmuir2020, Luo2022}. In printed systems, such effects provide a route to spatially graded chemical composition, filler concentration, and microstructure, enabling functional deposits with locally tailored properties~\cite{Kandemir2017,Xiaochun2026,Sutanto2012}. Nevertheless, how coalescence flows and subsequent drying jointly determine the final morphology remains less well understood.

A central unresolved question is therefore how strongly early coalescence dynamics imprint an orientational structure on suspended anisotropic particles, and to what extent this structure is preserved, weakened, or amplified during subsequent drying. This question is particularly important for anisotropic particles, as coalescence and evaporation-driven flows jointly determine the final orientational order and network connectivity of the deposited structure, thereby influencing its functional properties. In this work, we address this gap by studying how coalescence and subsequent drying generate dumbbell concentration and orientation distributions relevant to printed functionally graded materials. Given that the coalescence process is significantly faster than evaporation, we decouple these processes and organize the study into two complementary parts. First, we use asymmetric coalescence to determine how curvature and volume asymmetries generate directional flow, flow reversal, particle redistribution and alignment. Second, we use coalesced droplets as controlled reference states for evaporation, allowing us to isolate whether the orientational structure created during coalescence is preserved, enhanced, or weakened during drying. Together, these steps establish a physical framework for understanding how capillary coalescence and substrate-controlled drying jointly shape morphology in evaporating dumbbell-laden droplets.

\section{Methods}

We simulate the coalescence and evaporation of sessile droplets containing suspended dumbbells that represent general anisotropic particles. The liquid and surrounding gas are modeled as two immiscible fluid components using the lattice Boltzmann method, while the dumbbells are represented by bead--spring particles that are two-way coupled to the fluid. Over the past three decades, the lattice Boltzmann method (LBM) has emerged as a robust and versatile numerical technique for simulating multiphase and multicomponent flows~\cite{Krger2017, Liu2016}. Previously, we successfully developed and validated LBM-based models to investigate droplet coalescence~\cite{Thomas2023,xie2026c} and evaporation~\cite{Hessling2017,xie2025}, as well as the dynamics (e.g., assembly) of particulate suspensions composed of particles with diverse shapes~\cite{XieHarting2021}.

\subsection{Color-gradient lattice Boltzmann model}

The two fluid components are modeled using a multi-component color-gradient lattice Boltzmann method~\cite{Gunstensen1991,Leclaire2017}. For each component $k$, the distribution functions $f_i^k$ evolve according to
\begin{equation}
    f_i^{k}(\mathbf{x}+\mathbf{c}_i\Delta t,t+\Delta t) = f_i^{k}(\mathbf{x},t) - \frac{\Delta t}{\tau_k} \left[f_i^{k}(\mathbf{x},t)-f_i^{k,eq}(\mathbf{x},t)\right],
    \label{eq:lbe_method}
\end{equation}
where $\mathbf{c}_i$ are the discrete lattice velocities in direction $i$ and $\tau_k$ is the relaxation time of component $k$. The equilibrium distribution is given by
\begin{equation}
    f_i^{k,eq} = w_i\rho_k \left[1 + \frac{\mathbf{c}_i\cdot\mathbf{u}}{c_s^2} + \frac{(\mathbf{c}_i\cdot\mathbf{u})^2}{2c_s^4} -\frac{\mathbf{u}^2}{2c_s^2}\right],
    \label{eq:equilibrium_method}
\end{equation}
with lattice weights $w_i$, lattice speed of sound $c_s$, component density $\rho_k$, and fluid velocity $\mathbf{u}$. The macroscopic density and momentum fields are obtained from
\begin{equation}
    \rho_k=\rho_0\sum_i f_i^k, \qquad \rho\mathbf{u} = \sum_k\sum_i f_i^k\mathbf{c}_i, \qquad \rho=\sum_k\rho_k .
    \label{eq:macroscopic_fields_method}
\end{equation}
All quantities are reported in lattice units, with $\Delta x=\Delta t=\rho_0=1$.

Phase segregation is enforced by the recoloring step of the color-gradient method. Here, the local interface is described by the color field
\begin{equation}
    C = \frac{\rho_1-\rho_2}{\rho_1+\rho_2}.
    \label{eq:color_field_method}
\end{equation}
Surface tension is incorporated through a continuum-surface-force formulation~\cite{Brackbill1992,Gu2023,Akai2018}. The interface normal and curvature are computed from gradients of $C$, and the corresponding capillary force is applied as a body force before collision. This formulation replaces the classical perturbation step and improves numerical stability by reducing spurious currents.

Evaporation is modeled in the reaction-limited regime, where the rate is controlled by interfacial kinetics rather than by vapor diffusion~\cite{Nath2025}. Following the non-equilibrium one-sided model~\cite{Larsson2023,MURISIC2011}, the local evaporative flux is written as
\begin{equation}
    J = \frac{J_0}{K+\tilde{h}},
    \label{eq:neos_evaporation}
\end{equation}
where $\tilde{h}$ is the local droplet height normalized by the initial height, $J_0=0.0005$ sets the characteristic evaporation rate, and $K=10$ is the dimensionless kinetic resistance adopted from previous models of evaporating water droplets~\cite{MURISIC2011}. This expression follows from the Hertz--Knudsen relation after expressing the interfacial temperature in terms of the local film height under the thin-droplet approximation. Evaporation is applied at liquid--gas interface nodes as a mass sink that converts liquid resting populations into gas populations~\cite{Nath2025}. The evaporation rate is chosen sufficiently small so that the removed mass remains below the locally available liquid resting population, thereby avoiding unphysical negative populations during drying.

Wetting at the solid substrate is imposed using the geometric wetting boundary condition of Akai \textit{et al.}~\cite{Akai2018}. In this approach, wall normals are reconstructed from the voxelized substrate, and the direction of the color field gradient is modified at the wall boundary according to a specified contact angle. Contact-line motion is controlled by assigning different equilibrium contact angles to different regions of the substrate. In the present geometry, the lower contact angle $\theta_\mathrm{r}$ controls the receding part of the evaporating droplet, while the higher contact angle $\theta_\mathrm{a}$ acts mainly as a confining wetting condition that prevents spreading beyond the prescribed region. In the drying simulations below, $\theta_\mathrm{a}$ is kept fixed, whereas $\theta_\mathrm{r}$ is varied to tune the mobility of the receding contact line.

\subsection{Dumbell model and fluid coupling}

The suspended particles are represented as coarse-grained two-bead dumbbells. This minimal anisotropic model captures flow-induced redistribution and orientation while keeping the parameter space and computational cost manageable. Dumbbell models consisting of two connected beads are widely used as reduced representations of anisotropic particles and filament-like objects, particularly for studying their transport and alignment in external flows~\cite{Peters2007, Li2012}. The two beads of each dumbbell are connected by a finitely extensible nonlinear elastic (FENE) bond,
\begin{equation}
    U_{\mathrm{FENE}}(d) = -\frac{1}{2} k_s R_m^2 \ln\left(1-\frac{d^2}{R_m^2}\right), \qquad d<R_m ,
\end{equation}
where $d$ is the distance between the bead centers, $k_s=0.3$ is the spring constant, and $R_m=2.4$ is the maximum bond extension. Excluded-volume interactions between beads are described by a repulsive Weeks--Chandler--Anderson (WCA)-type potential,
\begin{equation}
    U_{\text{WCA}}(d) = \varepsilon \left[ \left( \frac{d_0}{d} \right)^{12} - 2 \left( \frac{d_0}{d} \right)^6 \right] + \varepsilon, \quad d \leq d_0,
\end{equation}
with bead diameter $d_0=2$ and interaction strength $\varepsilon=0.03$. Each dumbbell therefore has an effective length of approximately $4$ lattice units and an effective width of approximately $2$ lattice units. The initial dumbbell volume fraction is set to $5\%$ in all simulations. The dumbbells are a minimal coarse-grained representation of short anisotropic particles or filaments and are not intended as atomistic models of nanowires or nanotubes. They are initialized randomly inside the liquid phase. The suspension is initially dilute, so its feedback on the macroscopic capillary-driven flow generated during coalescence and evaporation remains weak. Local crowding can nevertheless develop during late-stage drying. The detailed parametrization follows our previous works~\cite{Schoettner2026,schoettner2026-2}.

The dumbbells are two-way coupled to the fluid by a dissipative drag force. Momentum exchanged between beads and fluid is back-coupled locally to the lattice following the approach of Ahlrichs and D\"unweg~\cite{Ahlrichs1999}. Confinement of the dumbbells inside the liquid phase is maintained by a solvation force based on local fluid-density gradients~\cite{Sega2013}. This force prevents the beads from leaving the droplet while allowing them to follow the internal flow. Particle--substrate interactions are modeled by a Lennard-Jones potential, following Ref.~\cite{Schoettner2026}, with the interaction strength denoted as $\varepsilon$. In the evaporation simulations, $\varepsilon$ is varied between $0.05$ and $0.2$ to test how substrate-induced immobilization affects the final dumbbells alignment.

\subsection{Simulation protocol and measured quantities}
\label{Sec:Protocol}

\begin{figure}
    \centering
    \includegraphics[width=1.0\linewidth]{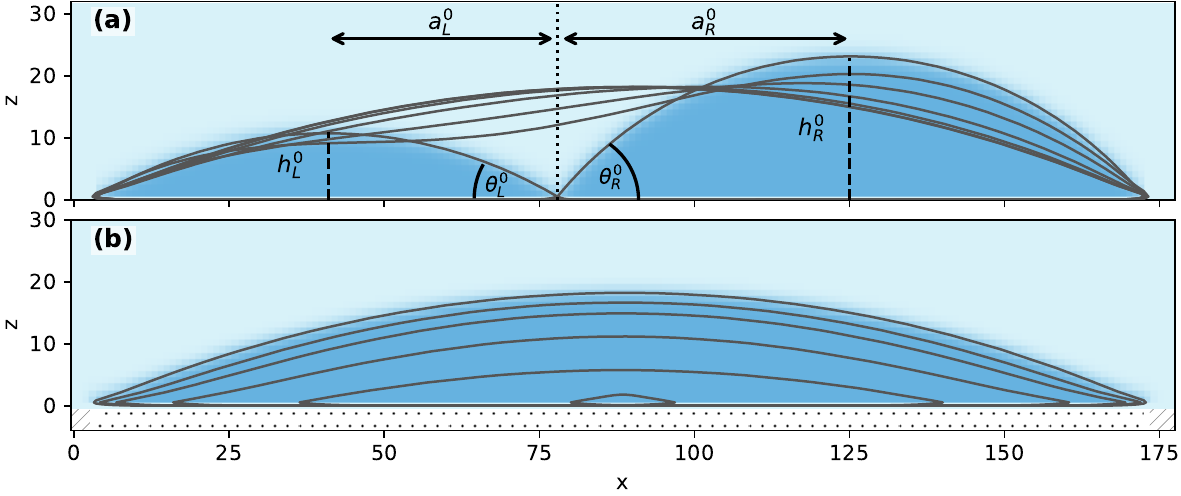}
    \caption{Side view of the temporal evolution of a quasi-two-dimensional liquid interface, indicated by gray contour lines. (a) Coalescence of two asymmetric droplets. (b) Subsequent evaporation of the merged droplet on a substrate with patterned wettability. The dotted region indicates the receding-contact-angle patch, while the outer dashed boundary marks the confining wetting patch.}
    \label{fig:Vis1}
\end{figure}

The simulations consist of two initially separated sessile droplets placed on a solid substrate, as illustrated in Fig.~\ref{fig:Vis1}. The geometry is quasi-two-dimensional, with each droplet represented as a cylindrical cap extended uniformly in the out-of-plane direction. The left and right droplets are characterized by their base radii $a_L^{0}$ and $a_R^{0}$ and contact angles $\theta_L^{0}$ and $\theta_R^{0}$, respectively. For each initial droplet, the corresponding radius of curvature is obtained from the circular-cap geometry as
\begin{equation}
    R_i^0 = \frac{a_i^0}{\sin\theta_i^0}, \qquad i\in\{L,R\}.
\end{equation}
The corresponding curvature is $\kappa_i^0=1/R_i^0$. For the coalescence analysis, geometric asymmetry is introduced by varying the base-radius ratio $a_L^{0}/a_R^{0}$, the contact-angle ratio $\theta_L^{0}/\theta_R^{0}$, or both, while keeping the mean initial base radius fixed at 42 lattice units and the mean initial contact angle fixed at $35^\circ$. Dumbbells are initially placed only inside the left droplet, so that their redistribution provides a direct measure of material transport induced by the coalescence flow. The study is divided into two parts. First, coalescence is simulated without evaporation to characterize the capillary-driven flow generated by bridge formation and subsequent shape relaxation. These simulations include asymmetric droplet pairs and are used to relate the final dumbbell alignment to the initial geometric asymmetry of the droplets. Second, evaporation is activated after the coalescence of two droplets with initially identical geometry. This controlled drying protocol is used to test whether the orientational order generated during coalescence is preserved, modified, or erased during evaporation, without introducing an additional asymmetry from the initial droplet shape. For the coalescence stage, time is nondimensionalized by the capillary--inertial time $t_c = \left(\frac{\rho \bar{R}^3}{\sigma}\right)^{1/2}$, where $\rho$ is the liquid density, $\sigma$ is the surface tension, and $\bar{R}=(R_L^0+R_R^0)/2$ is the mean initial radius of curvature of the two droplets. For the subsequent drying simulations, the two initially identical droplets are first allowed to coalesce without evaporation until $t=t_c$, by which time the dominant capillary-driven shape change has occurred. Evaporation is then activated from this state. This sequential protocol is used because coalescence usually occurs on a much shorter timescale than evaporation and because the prescribed evaporation profile is derived for a spherical-cap droplet and is therefore not directly applicable to the transient shape of coalescing droplets. The corresponding dimensionless coalescence time is $\tilde{t}=\frac{t}{t_c}$. Velocity fields during coalescence are scaled by the capillary--inertial velocity $ v_c = \frac{\bar{R}}{t_c} = \left(\frac{\sigma}{\rho \bar{R}}\right)^{1/2}$. Thus, the dimensionless velocity component in the coalescence direction is $ \tilde{v}_x=\frac{v_x}{v_c}$.

For the evaporation stage, time is normalized by the total evaporation time, $ t^* = \frac{t'}{t_{\mathrm{evap}}}$, where $t'$ denotes the elapsed time since evaporation is activated and $t_{\mathrm{evap}}$ is the total time required for the droplet to fully evaporate. Thus, $t^*=0$ corresponds to the beginning of evaporation and $t^*=1$ to the end of the drying simulation. The reported observables are averaged over independent simulation runs, denoted by $\langle \cdot \rangle$. The dumbbell orientation is quantified using a two-dimensional directional nematic alignment parameter along the coalescence direction $x$,
\begin{equation}
    S_x = 2\cos^2\zeta_{x,k} - 1,
\end{equation}
where $\zeta_{x,k}$ is the angle between the projected axis of dumbbell $k$ and the $x$-direction. This measure distinguishes alignment along the coalescence direction from transverse alignment within the substrate plane: $S_x=1$ indicates perfect alignment along $x$, whereas $S_x=-1$ corresponds to perfect transverse alignment within the substrate plane. The quantity $\left\langle \bar{S}_x \right\rangle$ denotes the run-averaged orientational order, where $\bar{S}_x$ is first averaged over all dumbbells in the domain. The value at the end of the coalescence stage is denoted by $\left\langle \bar{S}_x \right\rangle_c$. Spatial orientation profiles are obtained by sorting the dumbbell midpoints into bins along the $x$-direction. The corresponding binned orientational order is defined as
\begin{equation}
    S_x^b = \left\langle \frac{1}{N_b^*} \sum_{k=1}^{N_b^*} \left(2\cos^2\zeta_{x,k}-1\right)\right\rangle,
\end{equation}
where $N_b^*$ is the number of dumbbells in bin $b$. Spatial concentration profiles are computed analogously from the bead positions. For a bin of width $\Delta x_b$, the normalized bead concentration is defined as
\begin{equation}
    g^b = \frac{N_b/\Delta x_b}{N_0/(2a_L^0)}.
\end{equation}
Here, $N_b$ is the number of beads in the bin and $N_0$ is the total number of beads. With this normalization, $g^b=1$ corresponds to the mean bead density in the initially particle-containing left droplet. Values above and below unity, therefore, indicate local enrichment and depletion relative to this reference density. For dried deposits, positions are reported using the normalized coordinate
\begin{equation}
    x^*=\frac{x-x_L^0}{a_L^{0}},
\end{equation}
where $x_L^0$ is the center of the initially particle-containing left droplet. For the quasi-two-dimensional geometry considered here, the area of each initial droplet is written as
\begin{equation}
    \label{A_CAP}
    A_{\mathrm{cap},i} = \frac{{R_i^0}^2}{2} \left( 2\theta_i^0-\sin 2\theta_i^0 \right), \qquad i\in\{L,R\},
\end{equation}
where $A_{\mathrm{cap},i}$ is the circular-cap cross-sectional area. Since the system has the same depth in the out-of-plane direction for both droplets, the liquid volume is proportional to this area, $V_i \propto A_{\mathrm{cap},i}$. This relation is used to quantify the volume imbalance between the two droplets.

The simulations focus on the mesoscale mechanisms through which coalescence and evaporation affect dumbbell redistribution and alignment. The dynamics are governed primarily by surface tension, substrate wetting, and contact-line motion. Surface tension enters through the capillary pressure associated with interface curvature. This pressure scale is described by the Young--Laplace relation $\Delta p = \sigma \kappa$, where $\sigma=0.022$ is the surface tension in lattice units, and $\kappa$ is the interfacial curvature. Curvature differences during coalescence generate capillary-pressure gradients that drive the flow and relax the interface toward the merged circular-cap shape. During the subsequent evaporation stage, the interface remains close to this quasi-two-dimensional circular-cap geometry. The early coalescence dynamics are primarily capillary driven with moderate inertial effects. This is characterized by the Ohnesorge number $\mathrm{Oh}=\frac{\mu}{\sqrt{\rho \sigma \bar{R}}}$, where $\nu=1/6$ is the kinematic viscosity and $\mu=\rho\nu$ is the dynamic viscosity in lattice units. For the parameters considered here, $\mathrm{Oh}<1$, placing the system in the capillary--inertial regime. Evaporation occurs on a much longer timescale than coalescence, so the overall process can be interpreted as a rapid capillary rearrangement followed by slower evaporation-driven flow, contact-line motion, and particle deposition.

Microscopic contact-line mechanisms, such as molecular-scale slip or molecular-kinetic processes, are not resolved explicitly~\cite{Snoeijer2013,FernandezToledano2020}. Velocity-dependent dynamic contact-angle corrections are also neglected as the capillary number $\mathrm{Ca} =\frac{\mu U_{c}}{\sigma}$ is much smaller than unity, where $U_{c}$ denotes the characteristic contact-line velocity. Additional modifications of the wetting condition due to particle accumulation are not explicitly included~\cite{Weon2013}. Under these conditions, contact-line motion is assumed to be governed by the prescribed wetting pattern.

Evaporation is treated as isothermal, and the surface tension is kept constant. Thermocapillary and solutocapillary Marangoni stresses are therefore neglected. The carrier liquid is treated as Newtonian with constant viscosity. Concentration-dependent suspension rheology, including shear thinning or thickening induced by the dumbbells, is neglected, which is considered a reasonable approximation for the initially dilute suspension.

The dumbbells are modeled as dilute, weakly interacting suspended objects. Their inertia is neglected, corresponding to a small Stokes number $\mathrm{St} = \zeta_p / \zeta_f \ll 1 $, where $\zeta_p$ is the particle response time and $\zeta_f$ is a characteristic flow timescale. Thus, the dumbbells approximately follow the local flow and act mainly as probes of the accumulated flow-induced reorientation history. Their feedback on the macroscopic droplet shape is assumed to be weak. We note that although the initial suspension is dilute, evaporation-driven accumulation can locally increase the dumbbell concentration during drying. In particular, occasional dumbbell--dumbbell interactions may occur in the central deposition region, especially for high receding contact angles, where stronger contact-line confinement promotes particle accumulation.

The particles are represented by localized hydrodynamic coupling sites rather than by fully resolved, finite-size, anisotropic bodies. Dumbbell reorientation arises from the coupled motion of connected beads in the local flow field. Since the strongest deformation occurs during the finite coalescence and contact-line-motion stages, long-time periodic rotational dynamics are not expected to dominate the observed alignment.

The Bond and Galileo numbers are assumed to be small, so gravitational deformation and sedimentation are negligible compared with capillary forces. Brownian motion is also neglected because dumbbell transport and alignment are dominated by advective and capillary-driven flows on the length and time scales considered here. The simulations, therefore, focus on the deterministic coupling between capillary flow, substrate-controlled contact-line motion, evaporation, and the resulting dumbbell concentration and orientational order.

\section{Results}

We study the coalescence and evaporation of sessile droplet pairs on patterned substrates. Since the coalescence time scale is much shorter than the evaporation time scale under the present conditions, evaporation-induced flow is neglected during the initial coalescence stage. Moreover, the prescribed evaporation profile assumes a spherical-cap geometry and is not directly applicable during transient bridge formation and shape relaxation. Evaporation is therefore switched off during the rapid capillary-driven coalescence stage. Geometric asymmetry is introduced through the initial base radii and contact angles, while dumbbells are initially placed only in the left droplet to quantify flow-induced redistribution. In the second stage, evaporation is simulated after the coalescence of two droplets to examine how the coalescence-induced particle alignment is modified during drying and how the flow-induced structures are transferred into the final deposit morphology.

The initial droplet geometry is controlled by the base radii $a_L^{0}$ and $a_R^{0}$ and the contact angles $\theta_L^{0}$ and $\theta_R^{0}$. Fig.~\ref{fig:Vis1} shows a side view of the pure-liquid system. During coalescence, as shown in Fig.~\ref{fig:Vis1}a, a liquid bridge forms between the two droplets and grows until the merged droplet relaxes toward a combined circular-cap-like shape. During the subsequent evaporation stage, as depicted in Fig.~\ref{fig:Vis1}b, the contact line recedes according to the imposed wetting pattern. The dotted region corresponds to a receding contact angle e.g., $\theta_\mathrm{r}=20^\circ$, while the dashed region marks the confining outer patch associated with the confining wetting condition. After coalescence, the merged circular cap evaporates, and the contact line recedes inward when the apparent contact angle reaches the prescribed receding contact angle. For the symmetric drying cases considered here, this recession is symmetric with respect to the center of the merged droplet.

\subsection{Coalescence of pure droplets}

\begin{figure}
    \centering
    \includegraphics[width=0.55\linewidth]{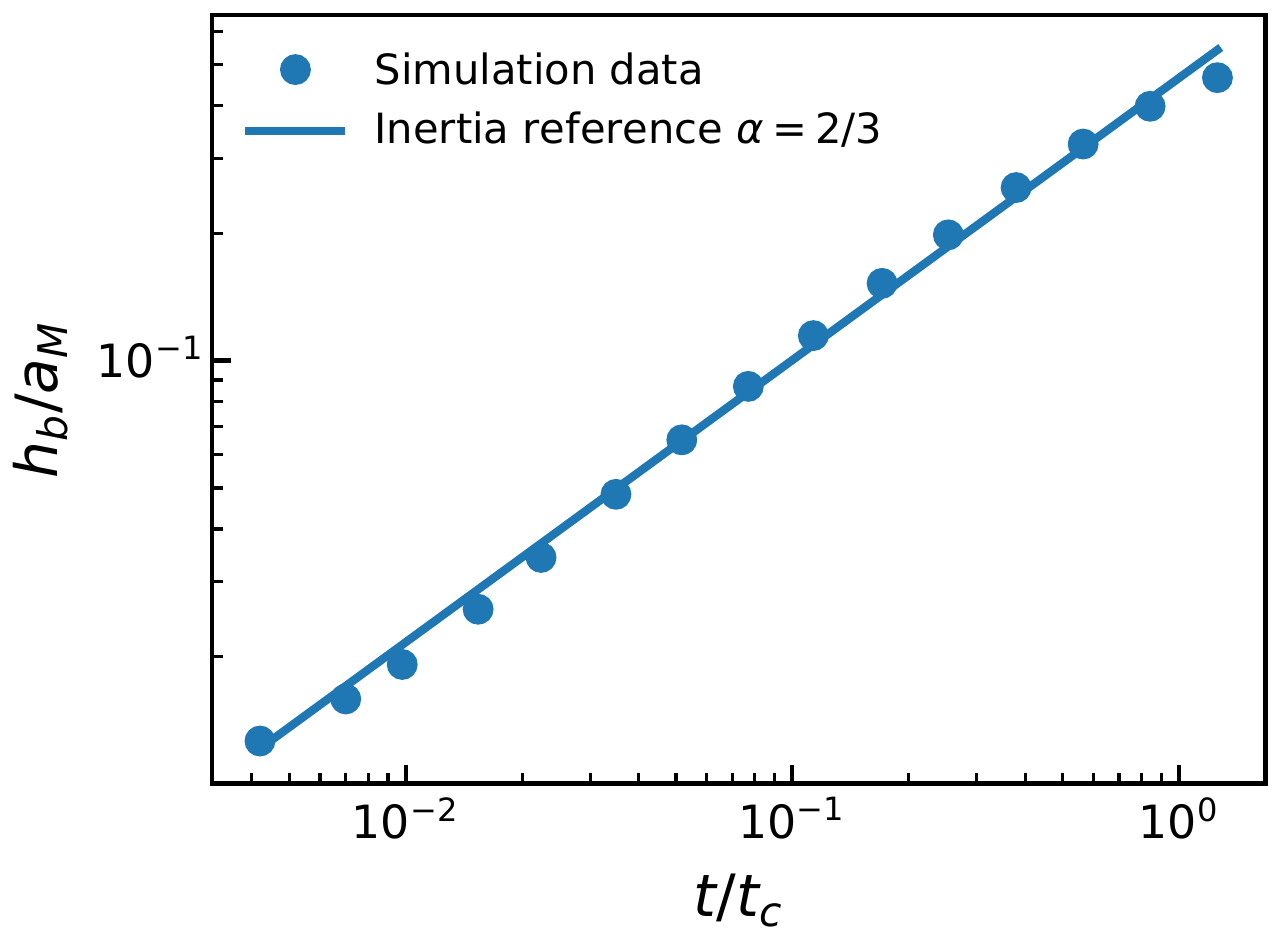}
    \caption{Bridge-height evolution during the coalescence of two asymmetric droplets. The initial droplets have the same base radius, $a_L^0=a_R^0=300$, and contact angles $\theta_L^0=90^\circ$ and $\theta_R^0=45^\circ$. The bridge height is normalized by the merged-droplet footprint radius $a_M$, and time is normalized by the capillary--inertial time $t_c=\sqrt{\rho \bar{R}^{3}/\sigma}$. The simulation data closely follow the reference scaling $h_b\propto t^{2/3}$ shown by the solid line, consistent with capillary--inertial coalescence dynamics.}
    \label{fig:Scaling}
\end{figure}

Before analyzing dumbbell alignment, we first characterize the pure-liquid coalescence dynamics that generate the transient flow field. The initial droplets have the same initial base radius, $a_L^0=a_R^0=300$, with contact angles $\theta_L^0=90^\circ$ and $\theta_R^0=45^\circ$. Fig.~\ref{fig:Scaling} shows the bridge-height evolution for representative asymmetric droplet pairs. The bridge height is normalized by the merged-droplet footprint radius $a_M$, and time is normalized by the capillary--inertial time $t_c=\sqrt{\rho \bar{R}^{3}/\sigma}$.  The bridge height follows a power-law growth with an exponent close to $2/3$, consistent with capillary--inertial coalescence dynamics~\cite{Pawar2019,Hack2020} in the present parameter range. This benchmark confirms that the early bridge growth follows the expected coalescence regime and provides the hydrodynamic background for the subsequent suspension analysis.

\begin{figure}
    \centering
    \includegraphics[width=1.0\linewidth]{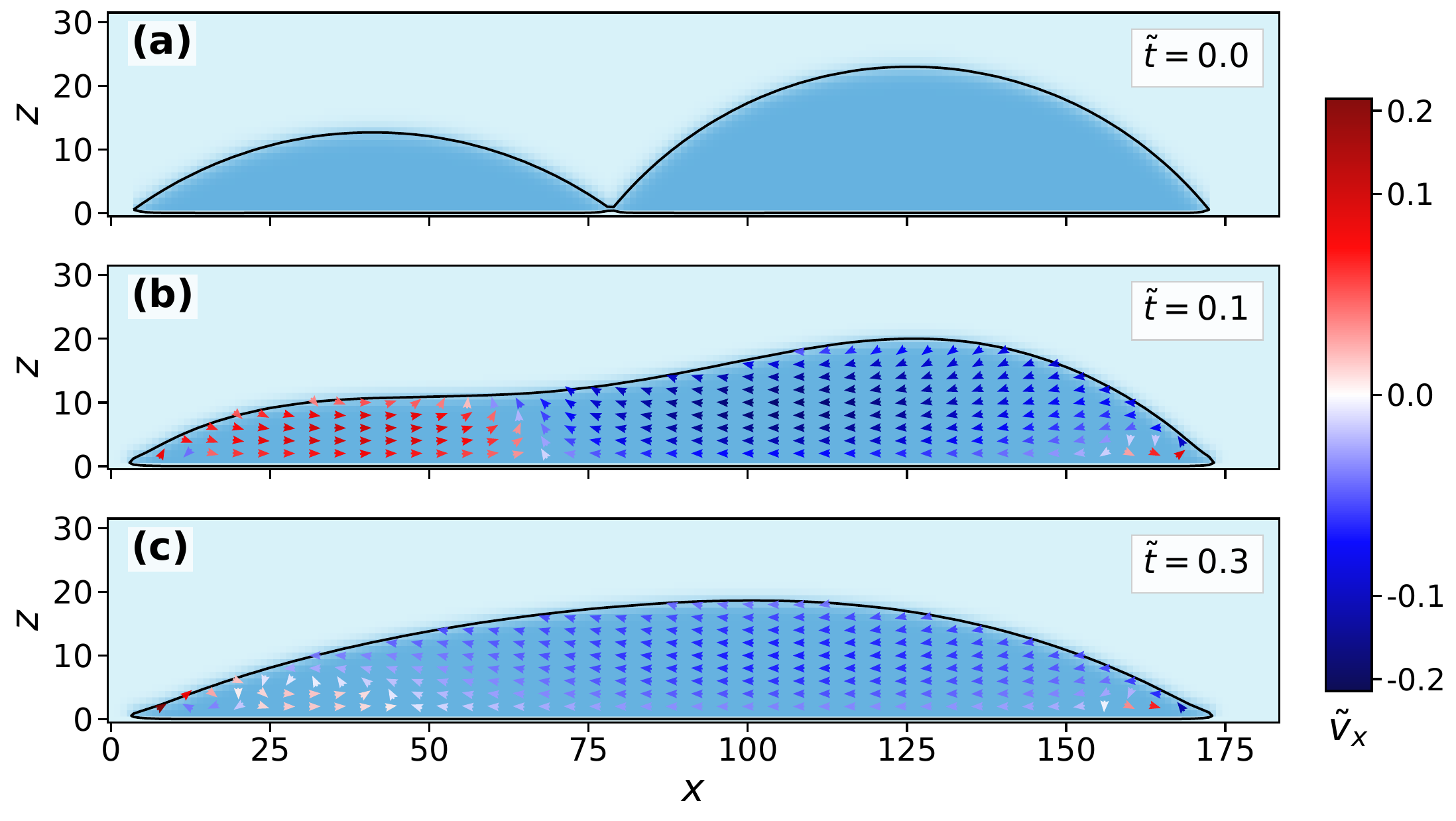}
    \caption{Snapshots of pure-liquid coalescence. The gray contour line indicates the liquid--gas interface, while the arrows show the velocity field. The colorbar shows the rescaled horizontal velocity $\tilde{v}_x$. Small spurious currents are visible near the contact lines at $x\approx 5$ and $x\approx 170$, but remain weak compared to the coalescence-induced flow. Their influence on the dumbbell dynamics is expected to be limited, because the dumbbells remain confined to the liquid phase and substrate friction restricts bead motion near the contact-line region.}
    \label{fig:Coal_Vis}
\end{figure}

Fig.~\ref{fig:Coal_Vis} shows the flow field generated during the coalescence of two asymmetric quasi-two-dimensional sessile droplets. The initial droplet geometry is shown in Fig.~\ref{fig:Coal_Vis}a. At early times, the flow is directed toward the newly formed liquid bridge, as shown in Fig.~\ref{fig:Coal_Vis}b. This motion is driven by the large capillary pressure associated with the highly curved bridge region, which generates an internal dynamic pressure during bridge growth~\cite{Pawar2019}. At later times (Fig.~\ref{fig:Coal_Vis}c), the flow field changes as the merged droplet relaxes toward its final capillary shape. In asymmetric cases, this relaxation can induce a redistribution of flow, because liquid must be transported between the two sides of the merged droplet. Thus, the coalescence flow is not purely bridge-directed, but contains a transient relaxation stage that can influence the orientation of suspended dumbbells.

\subsection{Coalescence-induced dumbbell alignment}

Elongated dumbbells are initially placed dilute in the left droplet and are used as orientational probes of the coalescence-induced flow. Their alignment is governed by the shear flow generated during bridge growth and droplet relaxation, rather than by the flow magnitude alone. For slender particles in flow, reorientation is governed by the local velocity-gradient tensor, and the orientational state of a suspension is commonly described by moments of the particle orientation distribution~\cite{Jeffery1922,AdvaniTucker1987,FolgarTucker1984}.

The base radii and contact angles of the two droplets are varied systematically. For each droplet-pair geometry, the results are averaged over six independent simulation runs with different random initial dumbbell configurations. Curvature asymmetry controls the capillary-pressure imbalance between the droplets, whereas volume asymmetry controls the amount of liquid redistributed during relaxation. These two geometric effects determine the strength and spatial structure of the coalescence-induced shear flow, and consequently govern dumbbell alignment.

\begin{figure}
    \centering
    \includegraphics[width=1.0\linewidth]{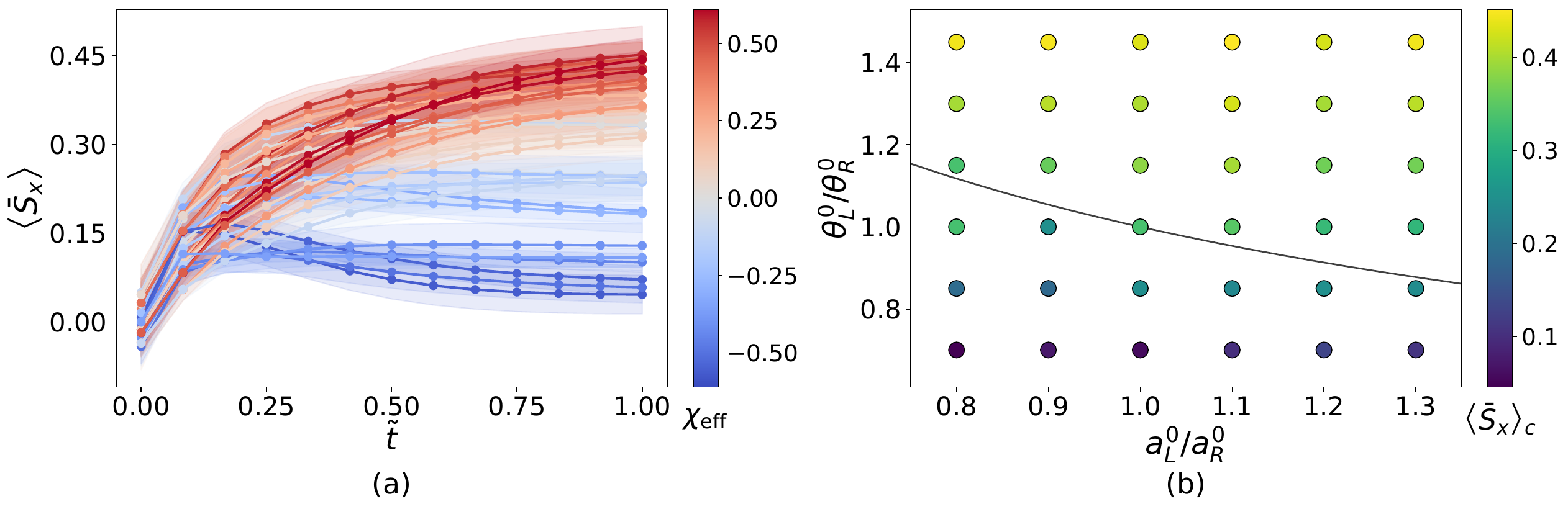}
    \caption{Nematic alignment induced by asymmetric droplet coalescence. (a) Temporal evolution of the mean nematic order $\langle \bar{S}_x\rangle$ during coalescence for different initial base radii and contact angles. The increase in $\langle \bar{S}_x\rangle$ shows that the coalescence flow aligns the dumbbells preferentially along the coalescence direction. The curves are colored by the effective geometric asymmetry parameter $\chi_{\mathrm{eff}}$, introduced in Sec.~\ref{sec:geometric_collapse}, and shaded bands indicate the standard error of the mean. (b) Final mean nematic order in the $x$-direction after coalescence for asymmetric droplets with different base-radius ratios $a_L^{0}/a_R^{0}$ and contact-angle ratios $\theta_L^{0}/\theta_R^{0}$. The solid line indicates the condition for zero net liquid redistribution during relaxation toward the final merged droplet shape.}
    \label{fig:nematic_comb}
\end{figure}

Fig.~\ref{fig:nematic_comb}a shows the temporal evolution of the mean nematic order in the $x$-direction as a function of the rescaled coalescence time $\tilde{t}$. For visual comparison, the curves are colored by the effective geometric asymmetry parameter $\chi_{\mathrm{eff}}$, which combines curvature and volume asymmetries and is introduced in the following Sec.~\ref{sec:geometric_collapse}. For all investigated parameter combinations, the nematic order in the $x$-direction increases during the early coalescence stage. This shows that the capillary-driven flow aligns the dumbbells preferentially along the main flow direction. Interestingly, for geometries with $\chi_{\mathrm{eff}}<0$ (dark blue lines), the nematic order can initially increase and subsequently decrease. Since $\langle \bar{S}_x\rangle$ is a nematic measure, this decrease cannot be caused by a mere reversal of the horizontal flow direction, because alignment along $+x$ and $-x$ contributes equivalently. Instead, the later relaxation stage likely exposes the dumbbells to a more complex strain field, including vertical flow components, recirculation and contact-line-induced reorientation. These contributions can rotate dumbbells away from the coalescence axis or broaden the orientation distribution, thereby partially weakening the initially generated $x$-alignment.

Fig.~\ref{fig:nematic_comb}b shows the final mean nematic order in the $x$-direction for different initial droplet asymmetries. The asymmetry is varied by changing the contact-angle ratio $\theta_L^{0}/\theta_R^{0}$ and the base-radius ratio $a_L^{0}/a_R^{0}$. The solid line marks the condition for effectively zero net liquid redistribution across the initial gap position, i.e., across the position where the liquid bridge first forms. To obtain this line, the initial areas $A_L^{0}$ and $A_R^{0}$ are computed from the circular-cap relation introduced in Eq.~\eqref{A_CAP}. The final merged droplet is then represented by a single circular cap with conserved total area. If $h_M(x)$ denotes the height profile of this final cap and $x_b$ the initial bridge position, the final area on the left side of the bridge is
\begin{equation}
    A_{M,L} = \int_{x_{\min}}^{x_b} h_M(x)\,\mathrm{d}x .
\end{equation}
The solid line is defined by the neutral redistribution condition $A_{M,L}-A_L^{0}=0$. Above this line, liquid is redistributed on average from the left, dumbbell-containing droplet to the right side of the final merged droplet, which supports the initial left-to-right motion and gives larger values of $\langle \bar{S}_x\rangle_c$. Below this line, the net redistribution occurs in the opposite direction after the first alignment has formed, which partly disturbs the dumbbell alignment along the $x$-direction and gives smaller final values of $\langle \bar{S}_x\rangle_c$. The final nematic order, therefore, reflects both the initial capillary driving and the subsequent redistribution needed to form the final merged-droplet shape.

\subsection{Geometric collapse of the final nematic order}
\label{sec:geometric_collapse}

To rationalize the dependence of the final nematic order on the initial droplet geometry, we compare the data using signed measures of curvature and volume asymmetry. The curvature asymmetry is directly related to the capillary-pressure imbalance,
\begin{equation}
    \Delta p \sim \sigma(\kappa_L^{0}-\kappa_R^{0}).
\end{equation}
Normalizing this pressure difference by the mean capillary-pressure scale $\sigma\bar{\kappa}$, with $\bar{\kappa}=(\kappa_L^{0}+\kappa_R^{0})/2$, gives
\begin{equation}
    \chi_\kappa = \frac{\Delta p}{\sigma\bar{\kappa}} \sim \frac{2(\kappa_L^{0}-\kappa_R^{0})}{\kappa_L^{0}+\kappa_R^{0}} .
    \label{eq:curvature_asymmetry}
\end{equation}
Here, $\chi_\kappa=0$ corresponds to equal initial curvatures, while the sign of $\chi_\kappa$ indicates which droplet has the larger curvature and therefore the larger Laplace pressure. The initial bridge-growth flow is directed toward the newly formed liquid bridge from both droplets. Thus, $\chi_\kappa$ does not determine the existence of this bridge-feeding flow itself, but quantifies the left--right imbalance of the capillary-pressure contribution relative to the symmetric case.

However, the same curvature imbalance does not necessarily lead to the same dumbbell alignment when the droplets have different volumes. The final orientation of the dumbbells reflects the total flow-induced strain experienced during coalescence, rather than only the instantaneous pressure difference at the onset of bridge formation. Curvature asymmetry mainly sets the capillary-pressure imbalance, whereas volume asymmetry determines how much liquid must be redistributed before the merged droplet relaxes toward its final shape. We therefore define the signed volume asymmetry
\begin{equation}
    \chi_V = \frac{V_L^{0}-V_R^{0}}{V_L^{0}+V_R^{0}},
    \label{eq:volume_asymmetry}
\end{equation}
where $V_L^{0}$ and $V_R^{0}$ are the initial liquid volumes of the left and right droplets. The condition $\chi_V=0$ corresponds to equal initial volumes, while the sign of $\chi_V$ indicates which droplet initially contains more liquid.

The relative magnitude of the curvature and volume contributions can be estimated from the first-order relation between curvature and volume asymmetry. For geometrically similar quasi-two-dimensional circular caps, the volume is proportional to the cross-sectional cap area and therefore scales as
\begin{equation}
    V \propto A_{\mathrm{cap}} \propto R^2 \propto \kappa^{-2}.
\end{equation}
To estimate the relative magnitude of the two asymmetry measures, we write the curvatures as small perturbations around the mean curvature,
\begin{equation}
    \kappa_i^{0} = \bar{\kappa}(1+\epsilon_i), \qquad |\epsilon_i|\ll 1 .
    \label{eq:small-per}
\end{equation}
Substitution of Eq.~\ref{eq:small-per} into Eq.~\ref{eq:curvature_asymmetry} gives
\begin{equation}
    \chi_\kappa \approx \frac{2(\epsilon_L-\epsilon_R)}{2+\epsilon_L+\epsilon_R}.
    \label{eq:curvature_asymmetry_expansion}
\end{equation}
Since $V\propto\kappa^{-2}$, the corresponding first-order volume scaling is
\begin{equation}
    V_i^{0} \propto (1+\epsilon_i)^{-2} \approx 1-2\epsilon_i .
\end{equation}
Thus, Eq.~\ref{eq:volume_asymmetry} becomes
\begin{equation}
    \chi_V \approx \frac{-(\epsilon_L-\epsilon_R)}{1-\epsilon_L-\epsilon_R}.
    \label{eq:volume_asymmetry_expansion}
\end{equation}
The ratio of the two asymmetry measures is therefore
\begin{equation}
    \frac{\chi_V}{\chi_\kappa} \approx - \frac{1}{2} \frac{2+\epsilon_L+\epsilon_R}{1-\epsilon_L-\epsilon_R} \approx -1 .
    \label{eq:volume_curvature_ratio}
\end{equation}
Thus, to first order, the signed volume asymmetry is approximately the negative of the signed curvature asymmetry,
\begin{equation}
    \chi_V \approx -\chi_\kappa .
\end{equation}
We assume that the flows induced by $\chi_\kappa$ and $\chi_V$ are of comparable strength and exert equal but opposite effects on dumbbell alignment. We therefore define the effective geometric asymmetry parameter as
\begin{equation}
    \chi_{\mathrm{eff}} = \chi_\kappa + \chi_V .
    \label{eq:effective_asymmetry}
\end{equation}

\begin{figure}
    \centering
    \includegraphics[width=0.7\linewidth]{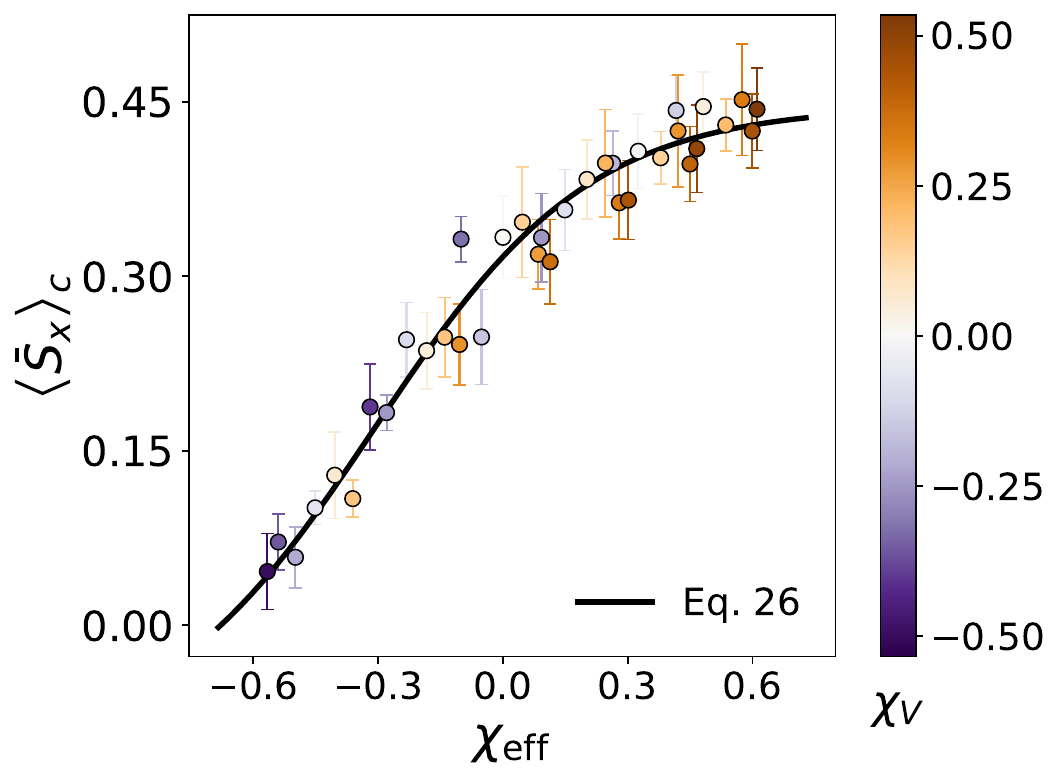}
    \caption{Collapse of the final mean nematic order in the $x$-direction as a function of the effective geometric asymmetry parameter $\chi_{\mathrm{eff}}$. Data obtained for different contact-angle and base-radius ratios collapse onto a common master curve. The solid line indicates the exponential fit given by Eq.~\ref{eq:tanh_nematic_collapse}. The color scale denotes the signed volume asymmetry $\chi_V$. Fit parameters are $S_0 = 0.17 \pm 0.05$, $S_{\mathrm{sat}} = 0.28 \pm 0.07$, $\chi_{\mathrm{c}}= -0.31 \pm 0.11$, and $w = 0.52 \pm 0.14$. Error bars indicate the standard error of the mean.}
    \label{fig:nematic_collapse}
\end{figure}

Fig.~\ref{fig:nematic_collapse} shows the final nematic order plotted as a function of $\chi_{\mathrm{eff}}$. Data obtained from different combinations of contact-angle and base-radius asymmetry collapse onto a common trend, indicating that the effective geometric asymmetry parameter $\chi_{\mathrm{eff}}$ provides a robust measure of coalescence-induced dumbbell alignment. The solid line in Fig.~\ref{fig:nematic_collapse} represents a fit based on an empirical exponential response function, given by
\begin{equation}
    \langle \bar{S}_x\rangle_c(\chi_{\mathrm{eff}}) = S_0 + S_{\mathrm{sat}}\tanh\left(\frac{\chi_{\mathrm{eff}} - \chi_{\mathrm{c}}}{w}\right),
    \label{eq:tanh_nematic_collapse}
\end{equation}
where $S_0$, $S_{\mathrm{sat}}$, $\chi_{\mathrm{c}}$ and $w$ are fitted constants. The hyperbolic tangent is chosen because the nematic order is bounded between $-1$ and $1$ and is expected to approach limiting values. The fit is used only as a compact master curve for the present data, not as a universal constitutive law. The collapse suggests that, within the explored parameter range, the final dumbbell alignment is governed primarily by the combined geometric parameter $\chi_{\mathrm{eff}}$, rather than by curvature or volume asymmetry independently.

\subsection{Evaporation of coalescence-aligned dumbbells}
\label{sec:evaporation_deposition}

We use the coalesced state obtained from two initially identical droplets as the starting configuration for evaporation. The initial droplets have base radii $a_L^{0}=a_R^{0}=42$ and contact angles $\theta_L^{0}=\theta_R^{0}=35^\circ$. Before coalescence, the dumbbells are placed only in the left droplet. Consequently, the finite nematic order and the inhomogeneous concentration profile at the beginning of evaporation result from the preceding coalescence process. During evaporation, we vary the receding contact angle $\theta_\mathrm{r}$ and the substrate-friction parameter $\varepsilon$ to determine whether the coalescence-induced alignment is preserved, disrupted, or restored during drying. The results are averaged over 60 independent realizations with different random initial dumbbell configurations.

\begin{figure}
    \centering
    \includegraphics[width=1.0\linewidth]{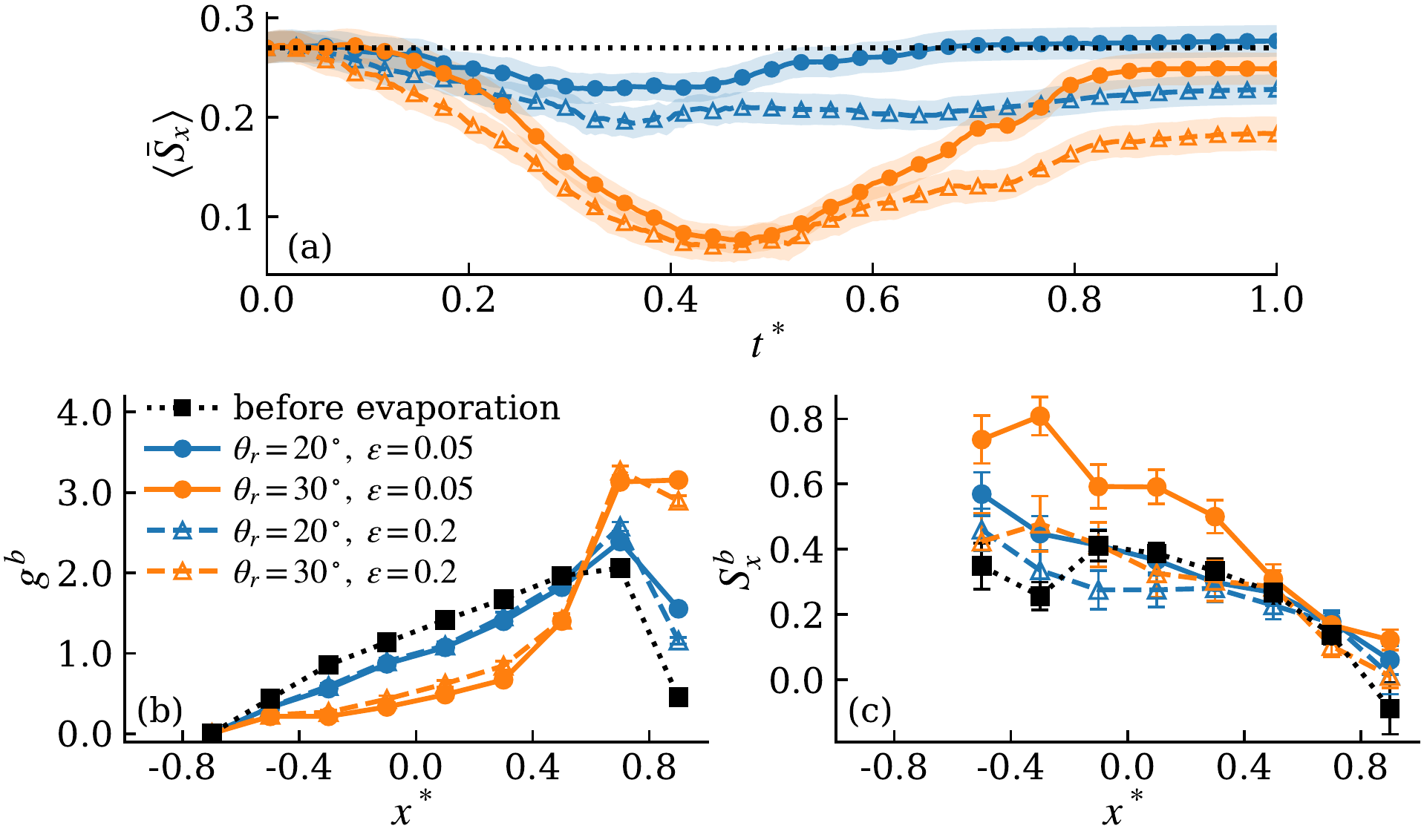}
    \caption{Temporal evolution of dumbbell alignment and spatial distributions of dumbbell concentration and alignment during evaporation for receding contact angles $\theta_\mathrm{r}=20^\circ$ and $30^\circ$ and substrate-friction parameters $\varepsilon=0.05$ and $0.2$. Colors distinguish the receding contact angles, while line styles and markers distinguish the substrate-friction parameters. (a) Temporal evolution of the global nematic order $\langle\bar{S}_x\rangle$ as a function of the normalized evaporation time $t^*$. The horizontal line indicates the mean nematic order immediately before evaporation. Shaded bands indicate the standard error across independent simulations. The bottom row shows spatial profiles along the coalescence direction $x^*$, centered at the initial center of the left droplet and normalized by its initial base radius $a_L^0$. (b) Final normalized dumbbell concentration profiles $g^b$ compared with the profile before evaporation. (c) Corresponding local nematic-order profiles $S_x^b$. Error bars in (b) and (c) indicate the standard error across independent simulations within each spatial bin.}
    \label{fig:COMB_evap}
\end{figure}

Fig.~\ref{fig:COMB_evap}(a) shows the temporal evolution of the global nematic order $\langle\bar{S}_x\rangle$. All cases start from the same mean pre-evaporation value, $\langle\bar{S}_x\rangle\approx0.27$, because the 60 independent simulations undergo the same coalescence stage before the receding contact angle and substrate-friction parameters are varied during the evaporation stage. During the first part of evaporation, the nematic order decreases for all investigated parameter combinations. This initial decrease reflects a reorganization of the dumbbells as evaporation changes the droplet geometry and transports the dumbbells along the resulting flow field. Changes in geometric confinement and the local streamline directions can rotate dumbbells away from the coalescence direction. Occasional dumbbell--dumbbell interactions may further disturb their orientation, although their contribution is expected to remain secondary due to the initially dilute dumbbell concentration. The temporary loss of alignment is considerably stronger for $\theta_\mathrm{r}=30^\circ$ than for $\theta_\mathrm{r}=20^\circ$. For $\theta_\mathrm{r}=30^\circ$, the global order reaches a minimum of approximately $0.07$ around $t^*\approx0.4$, whereas the decrease remains substantially smaller for $\theta_\mathrm{r}=20^\circ$. For $\theta_\mathrm{r}=20^\circ$, the higher substrate friction also produces a stronger decrease than the lower friction, showing that both parameters influence the intermediate orientational state.

After reaching its minimum, the nematic order increases again during the later stages of drying. The extent of this recovery depends strongly on the substrate friction. For the lower friction, $\varepsilon=0.05$, the case with $\theta_\mathrm{r}=20^\circ$ recovers to a value slightly above the pre-evaporation order, while the case with $\theta_\mathrm{r}=30^\circ$ reaches a slightly lower final value. At the higher friction, $\varepsilon=0.2$, the recovery is weaker for both contact angles, resulting in final values below the pre-evaporation reference. Higher friction limits dumbbell motion near the substrate and therefore reduces the ability of the dumbbells to rotate and recover their alignment during the later stages of evaporation. The receding contact angle mainly determines the overall strength of the temporary loss of alignment, while the substrate friction strongly affects the subsequent recovery.

The final concentration profiles in Fig.~\ref{fig:COMB_evap}(b) show that evaporation causes a pronounced redistribution of dumbbells along the coalescence direction. Here, $x^*=0$ corresponds to the initial center of the dumbbell-containing left droplet, while increasing $x^*$ points toward the initially dumbbell-free right droplet. The initial position of the bridge between the two droplets is located at $x^*=1$. Before evaporation, the concentration already increases toward positive $x^*$ as a result of dumbbell transport during coalescence. During drying, the concentration in the left part of the droplet decreases further, while the dumbbells accumulate in the bridge-side region at $x^*>0.7$.

The redistribution is stronger for $\theta_\mathrm{r}=30^\circ$ than for $\theta_\mathrm{r}=20^\circ$. With a contact angle of approximately $35^\circ$ at the beginning of evaporation, the receding threshold $\theta_\mathrm{r}=30^\circ$ is reached earlier than $\theta_\mathrm{r}=20^\circ$. The earlier contact-line recession transports more dumbbells toward the interior and the bridge-side region of the merged droplet. Consequently, the maximum normalized concentration exceeds $g^b=3$ for $\theta_\mathrm{r}=30^\circ$, compared with approximately $g^b=2.5$ for $\theta_\mathrm{r}=20^\circ$. The influence of substrate friction on the concentration profile is weaker than that of the receding contact angle, although the lower friction generally permits slightly stronger redistribution toward positive $x^*$.

Fig.~\ref{fig:COMB_evap}(c) shows that the final orientational order is also strongly position dependent. The local nematic order generally decreases from the left side of the original droplet toward the bridge-side accumulation region. The strongest local alignment is observed for $\theta_\mathrm{r}=30^\circ$ and $\varepsilon=0.05$, for which $S_x^b$ reaches values of approximately $0.8$ on the left side. Increasing the substrate friction reduces this local enhancement, consistent with a reduced ability of the dumbbells to rotate and realign during contact-line recession.

For all investigated parameter combinations, the local nematic order becomes small close to the initial bridge region, where the final dumbbell concentration is largest. The drying process therefore produces coupled spatial gradients in dumbbell concentration and orientation: the bridge-side region contains many weakly aligned dumbbells, while the left side contains fewer but more strongly aligned dumbbells. Spatial variations in composition and microstructure are central to functionally graded materials and can be used to obtain locally varying material properties~\cite{Saleh2020}. Such graded-material concepts have been investigated for a broad range of systems, including sensor and energy applications as well as soft material structures~\cite{Mueller2003,Akanksha2023}. The coupled concentration and orientation gradients observed here could therefore provide a route towards deposits with spatially varying properties. However, a direct calculation of the local conductivity would be required in future work to determine whether these structural gradients result in a corresponding functional gradient. The receding contact angle and substrate friction provide complementary control parameters: $\theta_r$ mainly determines where the dumbbells are deposited, whereas $\epsilon^*$ controls how long near-substrate dumbbells remain mobile enough to reorient before immobilization.

\section{Conclusion}
\label{sec:conclusion}

We investigated how coalescence and evaporation control the deposit morphology of drying sessile droplets containing dilute dumbbells and identified two coupled mechanisms that determine the final deposit structure. First, capillary-driven flow during bridge formation redistributes the suspended particles and induces preferential alignment along the coalescence direction. Second, evaporation-driven flow, contact-line motion, and particle--substrate friction further transport, concentrate, or immobilize the dumbbells during drying.

During coalescence, asymmetric initial droplet geometries generate a directional flow field at the early bridge growth stage. As the merged droplet relaxes toward a common capillary shape, liquid redistribution leads to a reversal of the dominant flow direction. In the presence of dumbbells, this transient flow history produces a measurable increase in the nematic order along the coalescence direction. For sufficiently asymmetric droplets, the later redistribution can partially weaken the initially generated alignment. The final nematic order can be rationalized by an effective geometric asymmetry that combines curvature and volume asymmetries, showing that both the capillary-pressure imbalance and the unequal liquid reservoirs contribute to the accumulated dumbbell alignment.

With evaporation, the global nematic order initially decreases as the evolving droplet geometry, contact-line motion, and internal flow reorganize the dumbbells. During the later stages of drying, the alignment can be partially recovered, but the extent of recovery depends strongly on substrate friction. Lower friction allows dumbbells to continue rotating and realigning, leading to a higher nematic order, whereas higher friction promotes earlier immobilization and results in a lower nematic order. The receding contact angle mainly controls the strength and timing of contact-line recession and therefore the spatial redistribution of dumbbells. Larger receding contact angles lead to stronger accumulation toward the bridge-side region, while smaller receding contact angles better preserve the coalescence-induced alignment.

Overall, our results show that coalescence and evaporation jointly determine concentration, alignment, and anisotropy in dumbbell-laden deposits. In particular, drying can transform a coalescence-preconditioned dumbbell distribution into coupled spatial gradients of concentration and orientational order. These insights provide design guidelines for controlling hydrodynamically generated microstructures in printed functionally graded materials.

Future work should extend this work toward quantitative experimental validation and establish direct links between the predicted deposit morphology and the resulting mechanical, optical, or electrical properties of the deposited material. The effects of larger contact angles, particle properties (e.g., length), and more viscous coalescence regimes should also be systematically investigated, as these factors may modify the coalescence-induced flow, particle reorientation, and the retention of alignment during drying. Extending the model to three dimensions and incorporating stronger particle--particle interactions at high concentrations will further elucidate the roles of out-of-plane reorientation, collective effects, and late-stage network formation in determining the final deposit morphology.

\vspace{1cm}

\subsection*{Author Contributions}
Conceptualization, J.S., Q.X. and J.H.;
methodology, J.S. and Q.X.;
software and validation, J.S.;
formal analysis and investigation, J.S. and Q.X;
data curation, J.S.;
writing---original draft preparation, J.S.;
writing---review and editing, J.S., Q.X. and J.H.;
visualization, J.S.;
supervision, Q.X. and J.H.;
project administration, J.H.;
resources and funding acquisition, J.H.

\subsection*{Funding}
This research was funded by the Deutsche Forschungsgemeinschaft (DFG, German Research Foundation) -- Project-ID 528402728 (research group ``3D-HF-MID"). We thank the Gauss Centre for Supercomputing e.V.(\url{www.gauss-centre.eu}) for funding this project by providing computing time through the John von Neumann Institute for Computing (NIC) on the GCS Supercomputer JUWELS at J\"ulich Supercomputing Centre (JSC).

\subsection*{Data Availability Statement}
The data that support the findings of this study are openly available at \url{https://doi.org/10.5281/zenodo.21710337}. 

\subsection*{Conflicts of Interest}
The authors have no conflicts to disclose.

\bibliography{Bibfile}
\end{document}